# Single-atom catalysts boost nitrogen electroreduction reaction

Yanling Zhai[a,b,†], Zhijun Zhu*[a,†], Chengzhou Zhu[c], Kyle Chen[d], Xueji Zhang*[b,e], Jing Tang*[f] and Jun Chen*[d]

[a]College of Chemistry and Chemical Engineering, Institute of Hybrid Materials, College of Materials Science and Engineering, Qingdao University, Qingdao 266071, P.R. China.

[b]Research Center for Intelligent and Wearable Technology, Qingdao University, Qingdao 266071, P.R. China.

[c]College of Chemistry, Central China Normal University, Wuhan 430079, P.R. China

[d]Department of Bioengineering, University of California, Los Angeles, CA 90095, USA.

[e]School of Biomedical Engineering, Shenzhen University Health Science Center, Shenzhen, Guangdong 518060, P.R. China.

[f]Department of Materials Science and Engineering, Stanford University, Stanford, CA 94305, USA.

[†]These authors contributed equally to this work.

*Correspondence: zhuzhijun@qdu.edu.cn (Z.Z.);

zhangxueji@szu.edu.cn (X.Z.);

jingtang@stanford.edu (J.T.);

jun.chen@ucla.edu (J.C.).

**Abstract**

Ammonia ($NH_3$) is mainly produced through the traditional Haber-Bosch process under harsh conditions with huge energy consumption and massive carbon dioxide ($CO_2$) emission. The nitrogen electroreduction reaction (NERR), as an energy-efficient and environment-friendly process of converting nitrogen ($N_2$) to $NH_3$ under ambient conditions, has been regarded as a promising alternative to the Haber-Bosch process and has received enormous interest in recent years. Although some exciting progress has been made, considerable scientific and technical challenges still exist in improving the $NH_3$ yield rate and Faradic efficiency, understanding the mechanism of the reaction and promoting the wide commercialization of NERR. Single-atom catalysts (SACs) have emerged as promising catalysts because of its atomically dispersed activity sites and maximized atom efficiency, unsaturated coordination environment, and its unique electronic structure, which could significantly improve the rate of reaction and yield rate of $NH_3$. In this review we briefly introduce the unique structural and electronic features of SACs, which contributes to comprehensively understand the reaction mechanism owing to their structural simplicity and diversity, and in turn expedite the rational design of fantastic catalysts at the atomic scale. Then, we summarize the most recent experimental and computational efforts on developing novel SACs with excellent NERR performance, including precious metal-, nonprecious metal- and nonmetal-based SACs. Finally, we present challenges and perspectives of SACs on NERR, as well as some potential means for advanced NERR catalyst.

# 1. Introduction

Ammonia ($NH_3$), one of the most important raw materials used in agriculture and the chemical industry, is mainly synthesized from pure hydrogen gas ($H_2$) and nitrogen ($N_2$) through the traditional Haber-Bosch process conducted under extremely harsh conditions (~40 MPa, ~500 °C), which consumes nearly 2% of the global energy supply.[1,2] In traditional $NH_3$ production, about 75% of energy input is used for $H_2$ generation from steam reforming and the remaining energy is applied to the Haber-Bosch process, which involves gas compression, $NH_3$ synthesis and post-separation. The resulting carbon dioxide ($CO_2$) emission from this process contributes to 1% of global greenhouse gases. In view of fossil fuel scarcity and global climate change, an energy-efficient and environment-friendly process for converting $N_2$ into $NH_3$ under ambient conditions is highly desirable. As an alternative approach to the Haber-Bosch process, electrocatalytic synthesis of $NH_3$ directly from $N_2$ and $H_2O$ has attracted considerable attention because the electrochemical process can be conducted under ambient conditions (without $H_2$ supply) and powered by renewable energy sources (Fig. 1).[3] Benefiting from the advantages of $N_2$ electroreduction reaction (NERR) such as its mild reaction condition, high energy efficiency, and functionality without $CO_2$ emissions or fossil fuel consumption, exploring smart design to synthesize efficient catalysts toward NERR has attracted growing interest.[4,5]

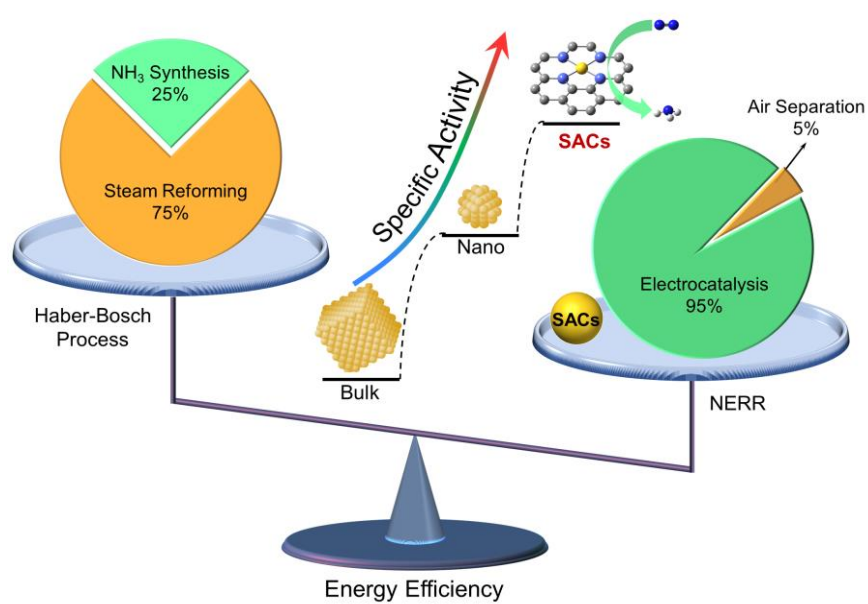

**FIGURE 1**. Schematic illustrating the energy-efficiency analysis between the conventional Haber-Bosch approach and the electrochemical approach to show the advantage of SACs over bulk and nanocatalysts.

The overall NERR reaction can be depicted as $N_2 + 3H_2O \rightarrow 2NH_3 + 3/2\ O_2$,[6] in which $N_2$ reacts with $H_2O$ to produce $NH_3$ and $O_2$. NERR has a similar theoretical potential to that of hydrogen evolution reaction (HER),[7] which creates a fundamental problem for electrochemical ammonia synthesis. NERR is typically plagued by the low affinity of $N_2$ on the catalysts, fast reaction kinetics of the competing HER,[8] much higher concentration of H than $N_2$ in aqueous solution and high bond energy of N≡N (941 kJ/mol), all of which always leads to a low $NH_3$ yield rate and a poor Faradaic efficiency (FE).[9-12] Particularly, HER always becomes the dominant process under highly negative potentials. Therefore, increasing efforts have been devoted to exploring catalysts with efficient NERR performance to improve the selectivity and reaction rate.

The size of nanostructures shows a significant influence on their NERR performance. As reported by Shi *et al.*, Au nanoclusters with a diameter of 0.5 nm supported on $TiO_2$ exhibited better NERR performance than other sizes (37 and 4 nm) regarding $NH_3$ yield rate, FE and selectivity, which was ascribed to the subnanometer size and the high dispersion of the Au nanocluster.[13] It is reasonable to expect that continuously minimizing the size of monodispersed Au nanoclusters on the substrate will lead to much improved catalytic performance because the atomically dispersed catalysts will maximize the interface of every single atom.[14] In addition, decreasing the size of catalysts will lead to an increased fraction of exposed atoms with unsaturated coordination sites, which have been proved to enhance the activity of electrocatalysts. Therefore, single atoms stabilized by supports would be promising candidates for NERR catalysts because of their maximized atom efficiency, unsaturated metal coordination environment, and unique electronic structures.[8]

Great efforts have been devoted to exploring single-atom catalysts (SACs) for NERR for following significant reasons. 1) SACs possess a similar structure to molecular catalysts showing high activity toward $N_2$ reduction, in which metal centers are coordinated by ligands. 2) Numerous couples of different atomic centers and various substrates would extend the sample capacity for mechanism investigation of NERR.[15] Although some excellent review articles have been published about the synthesis of novel catalysts and their applications in NERR [5,16-21] few of them have paid special attention to SACs.[22] Inspired by the great advantages of SACs and accumulated publications on SACs with high NERR performance, we summarize the most recent experimental and theoretical development of SACs for NERR under ambient conditions during the past few years. First, the unique structural and electronic features of SACs are briefly introduced. Furthermore, we focus on the most recent advances in the development of NERR SACs classified by the type of active centers. Finally, the major challenges and perspectives for future research on the design of efficient SACs for NERR and optimization of their electrocatalytic performance are presented. We expect that this review could inspire further interest in investigating the synthesis of novel NERR SACs, thereby offering opportunities in this emerging research area.

## 2. Single-atom catalysts

SACs are a kind of emerging catalyst composed of isolated single atoms (both metallic and nonmetallic ones) stabilized on supports. SACs have attracted increasing attention over the past several years since they were first proposed by Zhang *et al.*,[23] because they exhibit superior catalytic performance to their conventional counterparts in a range of reactions. SACs are synthesized through various wet-chemistry approaches, including spatial confinement, defect engineering, and coordination design strategies.[24] Single atoms are usually trapped by the defects or vacancies of the supports. The support offers a unique local environment for single atoms through the strong interactions between single atoms and the supports, such as unsaturated coordination and unique electronic structure,[23] which leads to large exposed active sites and maximized atom-utilization efficiency, thus enhancing intrinsic catalytic activity [25] and the low cost of the electrocatalysts.[26] With the aid of theoretical calculations, SACs are employed to reveal the mechanism of a given reaction because of the structural simplicity and homogeneity of a given SAC, which in turn promotes a comprehensive understanding of the synergetic effect of single atoms and supports and expedites the rational design of fantastic catalysts at the atomic scale. Recently, SACs have emerged as very promising catalysts for a range of reactions, such as oxygen reduction reaction,[27,28] HER,[29] oxygen evolution reaction,[30] CO oxidation reaction,[23] $CO_2$ reduction reaction,[31] and so on.[32,33] Most recently, taking advantage of the outstanding features of SACs and the unsatisfying performance of regular catalysts in NERR, researchers have turned their attention to SAC-based NERR catalysis.

## 3. Pre-screening of NERR SACs

Developing efficient, economical, and durable electrocatalysts is essential and significant for the rapid development and commercial application of NERR technology. Interactions between single atoms and supports are critical to the electrocatalytic activity and stability of SACs because there is a trade-off between diffusion and aggregation of single atoms, especially metal ones.[34] N-containing substrates, such as $C_3N_4$, N-doped graphene (NG), N-doped carbon (NC), N-doped porous carbon (NPC), zeolitic imidazolate frameworks (ZIF) and metal oxide with defects have been explored to stabilize single metal atoms because of their open porous structure, large surface area, abundant N species and plentiful metal and oxygen vacancies, which provides large possibilities to anchor and stabilize single atoms and mass transport during the electrochemical reaction.[35] In spite of the progress that has been made, it remains a great challenge to synthesize the ideal SAC for NERR due to the lack of an appropriate synthesis method. Alternately, density functional theory (DFT) calculations contribute to the evaluation of the activity of single atoms stabilized by the supports with a well-defined structure. To screen the NERR catalysts among the NG supported single metal atoms (Fig. 2A-E) more efficiently, Wang *et al.* proposed a general two-step strategy-based calculation (Fig. 2A) instead of determining all the reaction intermediates. The two-step strategy involves (1) $N_2$ capture and subsequent hydrogenation to $N_2H$ and (2) $NH_3$ generation from $NH_2$ and subsequent desorption. The two types of $N_2$ adsorption configurations (side-on and end-on), nine potential metal-NG coordination possibilities (Fig. 2B), and 30 kinds of metals (Fig. 2C) result in 540 possibilities. The $N_2$ adsorption energies ($\Delta E_{N_2}$) of different catalysts are shown in Fig. 2D. $\Delta E_{N_2}$ is highly related to the group number, the number of unoccupied d orbitals, and the d electrons of metals, as well as the coordination structures between metals and graphene substrates. After calculating the adsorption energy from evaluating the two key factors in each step, 10 promising candidates (Fig. 2E) with excellent NERR performances were selected. Among them, single W atoms with three C coordination ($W_1C_3$) afforded the best NERR activity. In addition to providing a series of promising SACs for NERR, this work proposed a simplified approach for the screening the catalysts toward NERR.

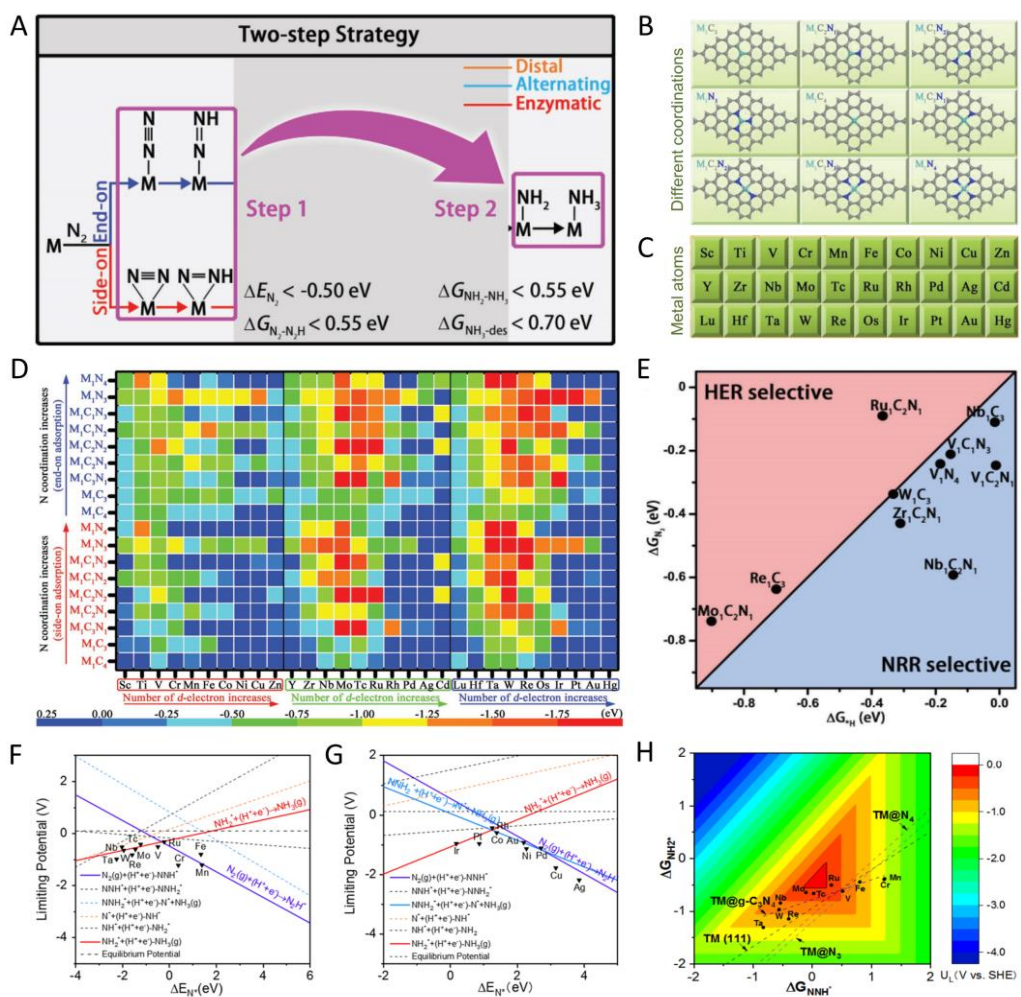

**FIGURE 2.** Calculation-based screening of NERR SACs. (A) Diagram of the two-step screening method. Proposed structures of NG-supported SACs with different coordination (B) and 30 kinds of transition metals (C). (D) Adsorption energies of $N_2$ on different SACs through side-on and end-on configurations. (E) Calculated adsorption free energies of H and $N_2$ on the selected SACs. Reproduced with permission.[36] Copyright 2018, Elsevier Ltd. (F-G) The relationship between limiting potential and the adsorption energy of *N ($\Delta E_{N^*}$) at 0 V for early (F) and late (G) TMs. * represents an adsorption state. (H) Color-filled contour plots of the limiting potential as a function of the Gibbs free energy of the key intermediates (*NNH and *$NH_2$) on the SACs with different compositions (metal centers and supports). Reproduced with permission.[15] Copyright 2019, American Chemical Society.

Most reported NERR catalysts suffer from low activity and poor FE, which greatly limited the development of NERR. A deep understanding of the mechanism could guide the rational design of NERR catalysts. Most recently, Qiao *et al.* have carried out DFT simulations to build a profile of potential SACs for NERR in consideration of the activity trends, electronic origins, and design strategies of the SACs.[15] The intrinsic activity of SACs was found to be highly related to the adsorption energy of N ($\Delta E_{N*}$, Fig. 2F and G). The variations of $\Delta E_{N*}$ on transition metals (TMs) originate from the bonding/antibonding orbital populations. The supports show great influence on NERR activity by introducing various ligand effects. The influence of electronic environments of 20 TMs on the activity of SACs was evaluated by using three kinds of NC substrates: g-$C_3N_4$ and carbons with different coordinating N atoms in NC. The limiting potential of SACs highly depends on the adsorption strength of key intermediates (Fig. 2H), because of the variation of the electronic structure of the SACs. Consequently, the suitable match between metal center and the supports, as well as the subsequent regulation of adsorption strength of the key intermediates are critical for the rational design of NERR SACs. In addition, stability and selectivity should be taken into accounts for catalyst design. As an example. single Ru and Rh atoms supported on g-$C_3N_4$ have been predicted as potential NERR catalysts over 60 candidates. This work made a significant contribution to the development of more efficient SACs for NERR.

## 4. Application of SACs in NERR

### 4.1 TM-based SCAs

TM-based catalysts show excellent activity to NERR because on the one hand, TM atoms have empty $d$ orbitals to accept the lone pair electrons of $N_2$, on the other hand, they are able to donate d electrons to the antibonding orbitals of $N_2$ and weaken the N≡N bond. Therefore, TMs with both empty and occupied d orbitals might be promising components for efficient NERR catalysts.[37,38]

#### 4.1.1 Precious metal-based SACs

Pt has been considered as a universal effective catalyst component for many reactions. However, Pt-based catalysts do not show appreciable activity for NERR in aqueous solution. First, Pt shows a much higher affinity for H than N. Second, the concentration of $H_2O$ is much higher than soluble $N_2$ in aqueous systems. Both of these two factors support the fact that Pt is prone to be covered with H instead of N. Thus a large overpotential is required to activate $N_2$ with Pt-based catalysts.[12,39] Fortunately, Ru- and Au-based SACs have been theoretically and experimentally proven to be efficient catalysts for NERR.

**4.1.1.1 Au SACs**

Electrocatalytic activities of metal catalysts are highly dependent on their electronic structures. Therefore, the state density of *d*-band metals is usually used to evaluate their electrocatalytic activities. Zhang *et al.* investigated the relationship between the coordination number and NERR activity of Au *via* first-principle calculations.[40] The NERR activity of Au is highly dependent on the coordination number of Au. Reduced coordination number of Au atoms would lead to exposed surfaces and enhanced electrocatalytic activity.[40] In addition, Au has been regarded to possess a low affinity to H, evidenced by the poor HER activity.[13] These intrinsic features endow Au with significant NERR activity. Qin *et al.* reported single Au atoms anchored on NPC (Au SA/NPC) for NERR (Fig. 3A).[7] The single Au atoms are stabilized by abundant N and C species with strong electronegativity, resulting in charge redistribution between Au atoms and the substrate.[41] Therefore, $N_2$ can be bridged between positively polarized Au atoms and negatively polarized N or C atoms, which facilitates the subsequent activation of $N_2$ by electronic polarization and further breakage of the N≡N bond. Consequently, the as-synthesized Au SA/NPC exhibited superior activity to NERR against HER, evidenced by the large current density difference over a wide potential range (Fig. 3B). To better understand the specific advantages of single Au sites, AuNPs anchored on NPC (AuNP-NPC) were used as a control. At a potential of -0.2 V (vs. RHE), Au SA/NPC revealed a much higher FE of 12.3% than that of AuNP-NPC (5.7%, Fig. 3C). Moreover, AuNPs anchored on N-free C (AuNP/C) exhibited a decreased FE of 4.5% at the same condition (Fig. 3C). These results indicated that both NPC substrate and single Au atoms play crucial roles in excellent NERR performance.[7]

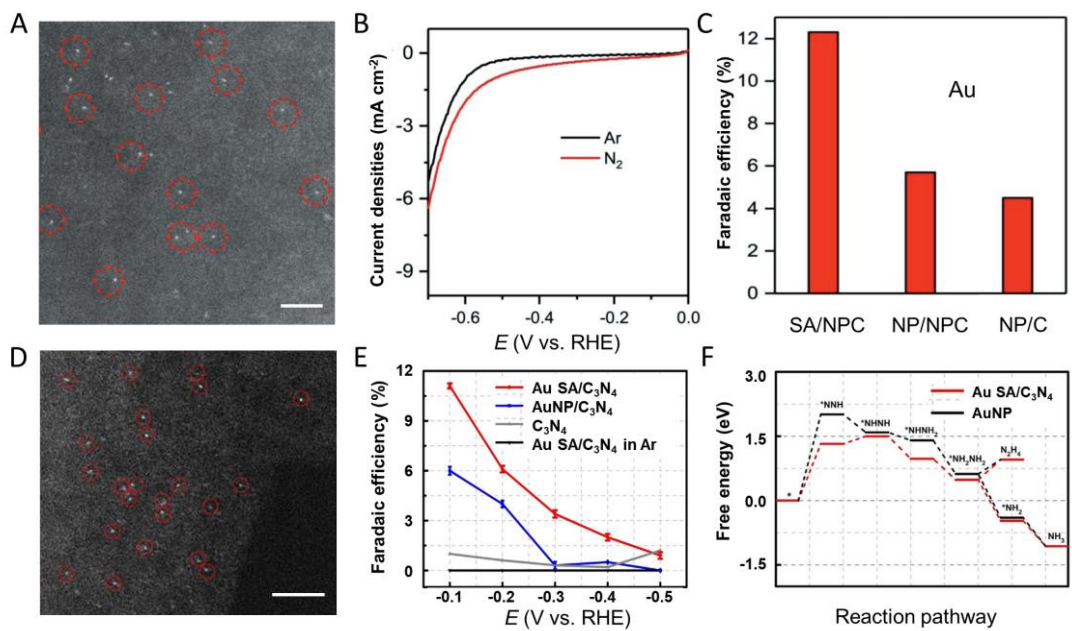

**FIGURE 3.** TEM images and electrocatalytic performance of the Au SACs. (A) High-magnification HAADF-STEM image of Au SA/NPC, scale bar: 2 nm. (B) LSV curves of Au SA/NPC in 0.1 M HCl aqueous solution saturated with $N_2$ and Ar. (C) FE values of $NH_3$ with different catalysts at -0.2 V (vs. RHE). Reproduced with permission.[7]. Copyright 2018, Wiley-VCH. (D) The aberration-corrected HAADF-STEM image of Au SA/$C_3N_4$ directly shows the atomic dispersion of Au atoms, scale bar: 2 nm. (E) FE values of $NH_4^+$ on different catalysts including Au SA/$C_3N_4$, AuNP/$C_3N_4$, pure $C_3N_4$, and Au SA/$C_3N_4$. (F) Free energy profile of NERR on Au SA/$C_3N_4$ and Au NP (211) at 0 V. Reproduced with permission.[42] Copyright 2018, Elsevier Ltd.

Both activity and selectivity are simultaneously required for an eligible catalyst. Wang *et al.* synthesized $C_3N_4$ supported single Au atoms (Au SA/$C_3N_4$) through $HAuCl_4$ adsorption and subsequent reduction under $H_2$ atmosphere. Au NPs anchored on $C_3N_4$ were synthesized for comparison. The bright dots circled in red in the aberration-corrected HAADF-STEM image (Fig. 3D) indicate the atomic distribution of Au. Au SA/$C_3N_4$ reached a high $NH_3$ FE than AuNP/$C_3N_4$ and $C_3N_4$ alone at a wide potential range (Fig. 3E). The free energy profiles shown in Fig. 3F reveals that the first hydrogenation step (reduction of $*N_2$ to $*NNH$) is the rate-determining step in NERR. The large free energy difference between the two catalysts of the first step explains the high NERR performance of SACs. In addition, compared to the formation of $NH_3$, the formation of $N_2H_4$ is energetically unfavorable for both Au SAC and AuNP. Benefiting from efficient atom utilization, Au SA/$C_3N_4$ displays excellent NERR efficiency with an FE of 11.1% and an $NH_4^+$ yield rate of 1,305 µg/h/mg$_{Au}$, which is 22 times higher than that of AuNP/$C_3N_4$.[42]

**4.1.1.2 Ru SACs**

Ru-based catalysts exhibit high activity in the traditional Haber-Bosch process[43] and have been considered as a new generation NERR catalyst with a calculated $N_2$ reduction overpotential lower than that of Fe.[10,44,45] Ru-based SACs have been extensively investigated for NERR from both theory and experiment. By using DFT computations, Liu *et al.* reported that single Ru atoms doped on electron-deficient B (Ru SA/B) exhibited outstanding NERR performance because of high $N_2$ affinity and the low energy barrier for formation of *$N_2$H and the destabilization of *$NH_2$ and $NH_3$ species. NERR took place on Ru SA/B *via* a distal pathway with an activation barrier of 0.42 eV, which was much lower than that of flat Ru catalysts (1.08 eV).[46] Recently, the electrocatalytic NERR performance of Ru SACs has been investigated experimentally. Geng *et al.* developed N-doped carbon-supported single Ru atoms (Ru SA/NC) catalysts by pyrolyzing Ru-containing ZIF (Fig. 4A). HRTEM image (Fig. 4B) and Extended X-Ray Absorption Fine Structure (EXAFS) spectrum confirm the atomic dispersion of Ru atoms stabilized by the surrounding N atoms serving as active centers for NERR. In contrast, Ru NP supported on NC (Ru NP/NC) was also synthesized by increasing the ratio of Ru/Zn in the precursors. The resultant Ru SA/NC catalyst exhibited an FE of 29.6% for NERR at -0.2 V (vs. RHE) in 0.05 M $H_2SO_4$, twice that of Ru NP/NC (Fig. 4D and E), and an $NH_3$ yield rate of 120.9 μg/h/$mg_{cat}$, which was an order of magnitude higher than the highest value ever reported. It also showed excellent durability with less than 7% decay of the $NH_3$ yield rate after a 12 h test. Moreover, the Ru SA/NC displayed approximated FE of $NH_3$ production in different electrolytes of various pH values, suggesting versatile applications of the catalyst. DFT calculations and temperature-programmed $N_2$ desorption results revealed that the excellent NERR performance of Ru SA/NC was attributed to the strong binding strength of $N_2$ on single Ru atoms. In addition, the Gibbs free energy change (ΔG) of *$N_2$ dissociation, the rate-limiting step, was much lower on Ru SA/NC than that on Ru NP/NC (Fig. 4C), leading to an enhanced NERR performance.[47]

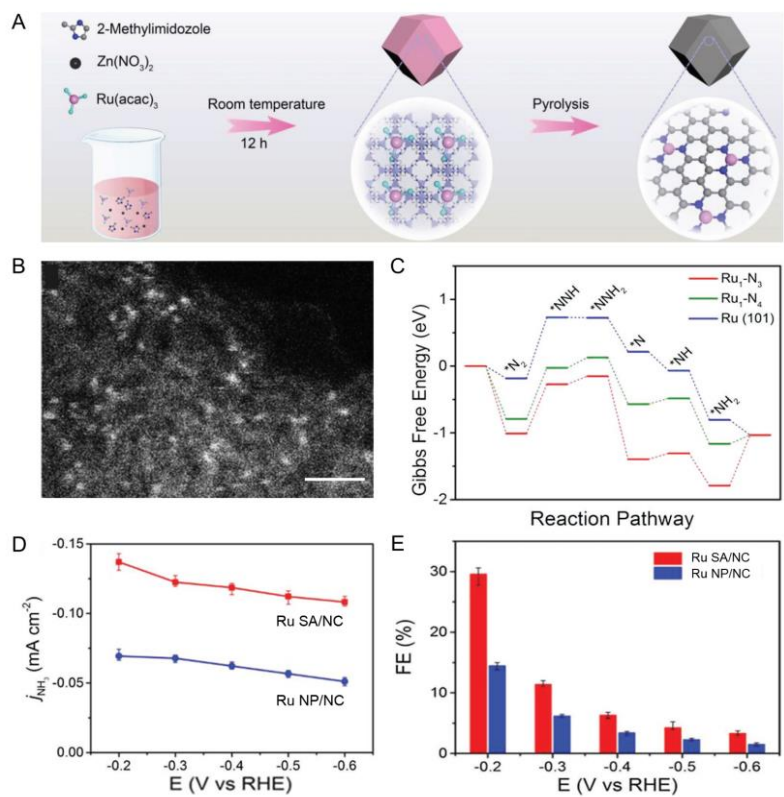

**FIGURE 4.** Preparation, characterization, calculation and performance of Ru SACs. (A) Schematic representation of the synthetic strategy for Ru SA/NPC. (B) HAADF-STEM image of Ru SAs/N-C (Scale bar: 1 nm). (C) Calculated free energy profile for NERR on various catalysts including $Ru_1-N_3$, $Ru_1-N_4$, and Ru (101) with a distal pathway. (D) Current densities and (E) FE of $NH_3$ production at different potentials on Ru SA/NPC and Ru NPs/NPC. Reproduced with permission.[47] Copyright 2018, Wiley-VCH.

It has been well-known that at an increased negative potential, the catalyst surface (especially for metal catalysts) would be thermodynamically covered by H adsorption (*H) rather than $N_2$, which not only facilities undesirable HER process, but also blocks the active sites for NERR, resulting in a significantly decreased FE.[48] Therefore, suppressing the H adsorption on the catalyst has been considered an efficient approach to improve FE and selectivity of the catalysts for NERR. Most recently, Tao *et al.* reported that the synergistic effect between single atoms and supports could promote their NERR performance.[10] NPC supported single Ru atom (Ru SA/NPC) synthesized through a coordination-assisted strategy exhibited high activity for $N_2$ electroreduction with an $NH_3$ yield rate of 3.665 mg/mg$_{Ru}$/h at -0.21 V, two times higher than that of the best catalysts ever reported. Interestingly, the addition of $ZrO_2$ was found to significantly promote NERR by suppressing the competitive HER. After the addition of $ZrO_2$, the single Ru atoms were observed to be embedded at $ZrO_2$ and NPC

(Ru@ZrO$_2$/NPC), since the O vacancies in ZrO$_2$ can bind Ru atoms and improve their stability. DFT calculations reveal that NERR mainly occurred at Ru atoms anchored at the O vacancies, which could stabilize the intermediate product, *NNH (*N$_2$ + (H$^+$ + e$^-$) →*NNH)), by reducing their formation free energy. Therefore, the initiation of the NERR process could be promoted because the formation of *NNH has been considered as the rate-limiting step in NERR on most metal catalysts.[49] ZrO$_2$ with O vacancy (Zr$_{32}$O$_{63}$) itself was inactive for NERR, however, Zr$_{32}$O$_{63}$ could significantly inhibit the competitive HER on Ru@Zr$_{32}$O$_{63}$, evidenced by much-reduced H adsorption free energy for Ru@Zr$_{32}$O$_{63}$. As a result, Ru@ZrO$_2$/NPC delivered a stable NERR behavior over 60 h and 21% of faradic efficiency at -0.11 V. The outstanding NERR performance of Ru SACs was attributed to the atomic dispersion of Ru, HER-suppression effect of ZrO$_2$, enhanced N$_2$ adsorption and low energy barrier for the first protonation of N$_2$.[10] In addition, other precious metal (such as Rh and Os)-based SACs have been predicted as potential catalysts for NERR.[50,51] However, most of them have not yet been experimentally examined.

**4.1.2 Non-precious metal-based SACs**

Despite their excellent NERR performance, wide application of precious metal-based catalysts has been restricted because of low abundance, high cost, and short catalyst life. Non-precious metal-based nanostructures have been considered as a candidate for NERR catalyst. Inspired by the fact that Fe- and Mo-based structures work as the active sites in enzyme nitrogenase, single Fe and Mo atom catalysts are demonstrated to be the most promising candidates for NERR from both experimental[8,52] and DFT calculations.[53,54]

### 4.1.2.1 Fe SACs

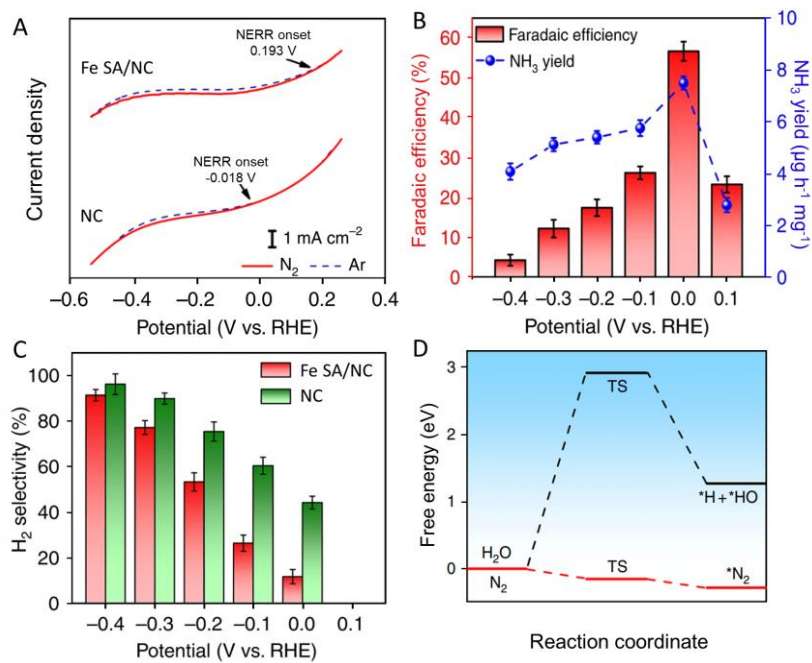

**FIGURE 5.** The NERR performance of Fe SA/NC and NC. (A) Linear sweep voltammetry (LSV) curves of Fe SA/NC and NC in 0.1 M KOH aqueous solution saturated with Ar and $N_2$ with a scan rate of 50 mV/s. (B) $NH_3$ FEs and yield rates of Fe SA/NC at different potentials. (C) $H_2$ selectivity profile of Fe SA/NC and NC at each given potential. (D) Calculated free energies of the adsorption between *H and *$N_2$ on Fe SA/NC. Reproduced with permission.[9] Copyright 2019 Nature Publishing Group.

Fe has been recognized as the active center in both the Haber-Bosch process and biological nitrogen fixation. With this in mind, Fe SACs are expected to be efficient catalysts for NERR. NERR is typically hindered by a high reaction barrier and competing HER, which leads to a low FE. Synthesizing the catalysts with a positive $N_2$ reduction potential is an efficient way to suppress HER because the theoretical potential of HER is 0 V. To this aim, Wang *et al.* synthesized single Fe atoms supported on NC (Fe SA/NC), which exhibited a positive-shifted onset potential of NERR at 0.193 V (Fig. 5A), which is much higher than that of NC (-0.018 V). Therefore, Fe SA/NC afforded a superior FE of 56.55% and a high $NH_3$ yield rate of 7.48 µg/h/mg at 0 V (Fig. 5B), because the HER is greatly suppressed under this potential. Compared with the NC substrate, Fe SA/NC is superior in terms of FE, yield rate and selectivity over the whole tested potentials (Fig. 5C). The decreased valence state of Fe in the initial NERR process facilitates $N_2$ binding to Fe sites by donating electrons from $N_2$ to the unoccupied *d* orbitals of Fe. DFT simulation results revealed that both *H (dissociated from water) adsorption and subsequent desorption to form $H_2$ requires high energy, indicating a significantly suppressed HER. On the other hand, $N_2$ adsorption on Fe sites was calculated to be an exothermic process (Fig. 5D), which favors the subsequent hydrogenation process and leads to excellent NERR performance with high NERR selectivity.[9]

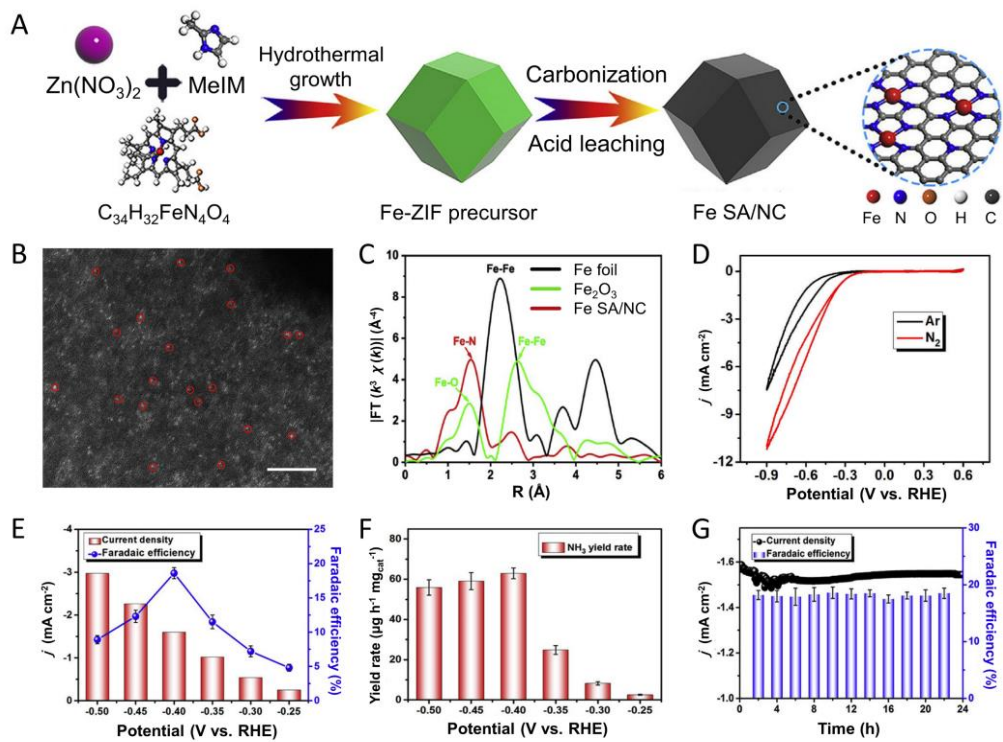

**FIGURE 6.** The preparation, characterization and NERR performance of Fe SACs. (A) Schematic synthesis of Fe SA/NC. (B) HAADF-STEM image of Fe/SA/NC, scale bar: 2 nm. (C) EXAFS profile of Fe SA/NC. (D) CV curves of Fe SA/NC in $N_2$- and Ar-saturated 0.1 M PBS solutions. (E) Electrocatalytic current densities and FE values of Fe SA/NC in 0.1 M PBS solution saturated with $N_2$ at given potentials. (F) $NH_3$ yield rate at different potentials. (G) NERR stability test of Fe SA/NC over 24 h. Reproduced with permission.[52] Copyright 2019, Elsevier Ltd.

Most reported SACs for NERR are working in acidic or alkaline electrolytes, and these corrosive electrolytes easily cause erosion of the substrate and destabilization of the single atoms. Neutral aqueous electrolytes are considered environment-friendly because it can reduce the corrosion issue and cut down the cost of the electrochemical system. For the first time, Liu *et al.* have developed neutral aqueous electrolyte-working Fe-based SACs for efficient NERR through hydrothermal and anneal treatments of the mixture of Fe/Zn-containing ZIF (Fig. 6A). Fe atoms are atomically dispersed on the NC (Fig. 6B). From the extended X-ray absorption fine structure (EXAFS) fitting analysis, it can be found that the active center, an atomically dispersed Fe atom, is coordinated by four surrounding N atoms (Fig. 6A), and the corresponding Fe-N bond distance was measured as 1.5 Å (Fig. 6C). The as-prepared Fe SA/NC exhibits a high selectivity toward NERR (Fig. 6D), an FE of 18.6% (Fig. 6E) and an $NH_3$ yield rate of 62.9 μg/h/$mg_{cat}$ (Fig. 6F), as well as high durability (slight decay over 24 h test) at -0.4 V in 0.1 M phosphate buffer solution (PBS, Fig. 6G). The excellent performance for NERR is ascribed to the high density of atomically dispersed Fe-$N_4$ species in the catalysts. DFT calculations reveal that the Fe atoms coordinated with N can donate electrons to adsorbed $N_2$ to elongate the N≡N bond length, thus facilitating the subsequent hydrogenation of the adsorbed $N_2$.[52]

**4.1.2.2 Mo SACs**

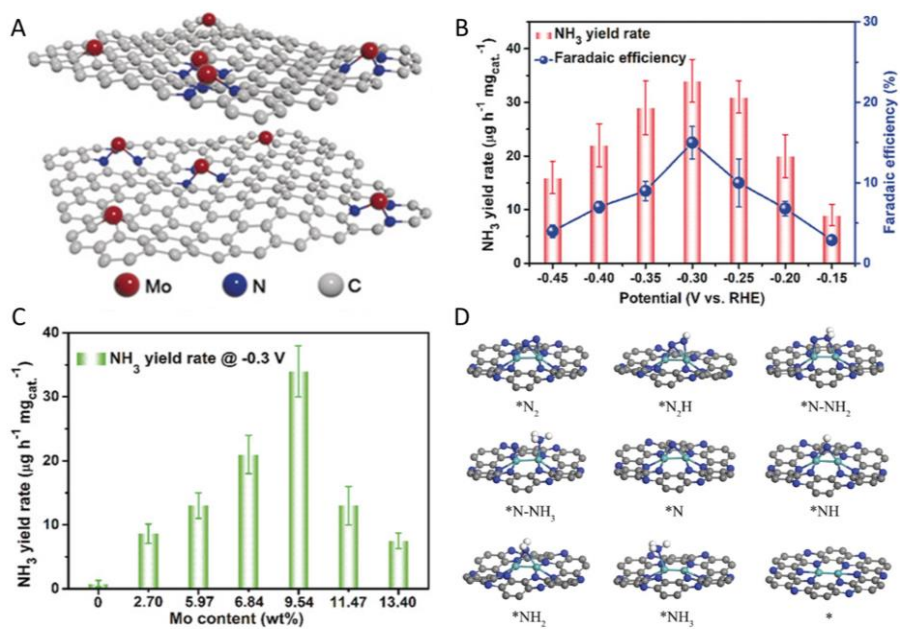

**FIGURE 7.** Proposed structures and catalytic activities of Mo SACs. (A) Atomic structure model of Mo-SA/NPC. (B) FE (blue) and $NH_3$ yield rate (red) at different potentials. (C) $NH_3$ yield rates of Mo-SA/NPC catalysts with different Mo loading amounts. Reproduced with permission.[8] Copyright 2019 Wiley-VCH. (D) Proposed configurations of the intermediate of each step on the active centers of $Mo_2$/$C_2N$. Reproduced with permission.[55] Copyright 2018 Royal Society of Chemistry.

Considering that the active centers of most natural nitrogenases are Mo-based structures, Mo SACs would be an efficient NERR catalyst. Han *et al.* synthesized single Mo atoms anchored on NPC through pyrolysis of a mixture of glucose, $(NH_4)_6Mo_7O_{24}$ and $NH_2OH·HCl$.[8] The single Mo atoms are stabilized *via* the formation of C-Mo and N-Mo bonds (Fig. 7A), confirmed by XPS and XANES results. Fig. 7B indicates that the as-prepared Mo SA/NPC catalyst with a high Mo loading of 9.54 wt% exhibits a high $NH_3$ yield rate of 34.0 μg/mg$_{cat}$/h (31.5 μg/mg$_{cat}$/h) and FE of 14.6% (6.8%) at -0.3 V (-0.25 V) in 0.1 M KOH (0.1 M HCl) which is attributed to the porous carbon structure and a large number of active sites (Mo-N) because Mo-N was reported to activate *$N_2$, stabilize *$N_2H$ and destabilize *$NH_2$ species.[53] However, Mo nanoclusters appeared after further increasing the Mo content, while the electrocatalytic performance declined with reduced amount of Mo-N species (Fig. 7C), which confirmed the critical role of single Mo atoms in NERR.[8] With an aid of first-principles computations, dual Mo atoms anchored $C_2N$ ($Mo_2/C_2N$) was predicted to show the best NERR performance among many other TM-based catalysts.[55] As shown in Fig. 7D, $N_2$ was firstly adsorbed on the $Mo_2N_6$ with a side-on configuration, and the N–N bond was elongated to 1.21 Å by the interaction between *$N_2$ and catalyst. Then the *$N_2$ would be hydrogenated by an adsorbed proton with a barrier of 0.31 eV, which led to an increased N–N bond of 1.33 Å. In turn, *NNH was hydrogenated much easier (0.25 eV) because of the decreased bond energy. Finally, $N_2$ was reduced in a distal way to release the first $NH_3$. The remaining N will be successively hydrogenated to produce the second $N_2$. The enhanced NERR performance was ascribed to the following factors: 1) N atoms with negative charge could fix Mo atoms firmly; 2) $Mo_2N_6$ moieties act as the active center to donate electrons to the antibonding orbital of $N_2$ to facilitate the activation of $N_2$; 3) the weak HER activity on $Mo_2/C_2N$.[55]

**4.1.2.3 Others**

Besides Fe and Mo, other metal-based SACs have been predicted as potential catalysts for NERR through computing studies. For example, single W atoms coordinated with three C atom embedded in graphene ($WC_3$) were selected as a promising catalyst through a high-throughput screening analysis among hundreds of TM-based SAC candidates.[36] Ni SA/$WS_2$ shows good energy-efficiency and good stability for NERR because it can accept electrons from *$N_2$ to waken the bonding energy of N≡N.[56] Zhao *et al.* identified single Mo/Cr atoms on NG (Mo SA/NG and Cr SA/NG) as a NERR catalyst based on calculations in terms of stability, competitive adsorption of $N_2$ and H, and formation potential of *NNH and *H on metal sites.[57]

Although Y- and Sc-based nanomaterials are inactive to most reactions at room temperature, Y and Sc single atoms with six coordination bonds (Y SA/NC or Sc SA/NC) have been reported to show catalytic functions toward NERR. DFT calculations reveal the proper adsorption and activation energies of $N_2$ on the Y SANC and Sc SA/NC, because of the unique local electronic structures of single Y/Sc atoms arised from the unusual coordination structure.[58] Co dimer anchored on graphdiyne ($Co_2$/graphdiyne) was evaluated as a good NERR catalyst among TM (Co, Ni, Fe, and Mn) monomer, dimer and trimer stabilized on graphdiyne monolayers because of the HER-suppressing effect of $Co_2$/graphdiyne and strong electron-donating ability of graphdiyne monolayer.[59] These investigations not only present a series of promising NERR catalysts, but also provide important guidance for the rational design of excellent catalysts for NERR.

**4.2 Metal-free SACs**

TM-based SACs deliver excellent NERR performances, however, there are still some intrinsic drawbacks: the d electrons of TMs facilitate the adsorption of $H^+$ and subsequent HER; TMs in SACs might be released in some harsh conditions.[60,61] Fortunately, metal-free catalysts exhibit weak H adsorption and better stability. It is worth mentioning that heteroatom doping could tailor the electronic structure of carbon materials due to the strong electronegativity of heteroatoms and consequent high positive charge density of the adjacent carbon atoms, which has been proved as an efficient way to boost their catalytic performance.[62,63] With this in mind, the catalytic performance of metal-free SACs for NERR has been explored.

### 4.2.1 B-SACs

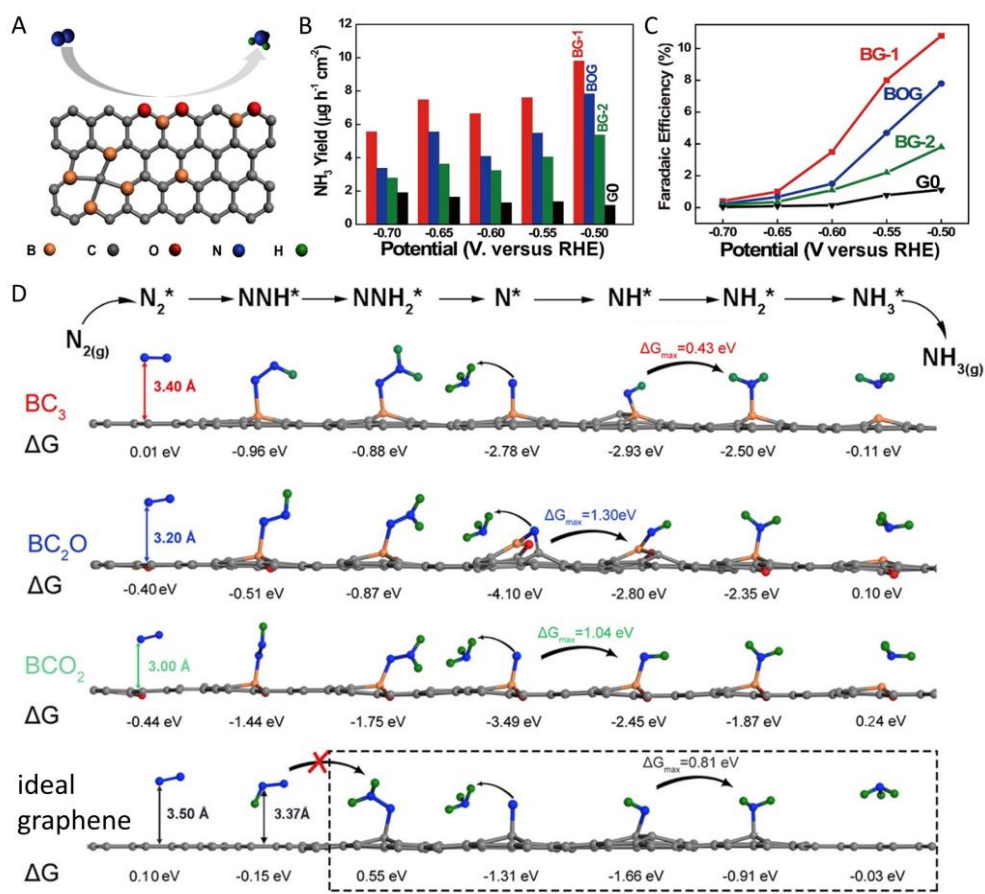

**FIGURE 8.** Experimental results and theoretical prediction of B doped G for NERR. Proposed B-doped G with different coordination types. The NH$_3$ production rates BG-1, BOG (same as BG-1 except thermal treatment under Ar), BG-2, and G at different potentials. FE of BG-1, BOG (same as BG-1 except thermal treatment under Ar), BG-2, and G at different potentials. (D) Reaction pathways and the corresponding energy changes of NERR on indicated catalysts. Reproduced with permission.[64] Copyright 2018, Cell Press.

Electron-deficient B can efficiently tune the electronic structure of carbon nanomaterials. For example, after being doped with B, graphene could retain its original sp$^2$ hybridization and planar structure. Because of the lower electronegativity of B than C, B atoms are ready to be positively charged in B doped carbon materials, which would promote NERR by enhancing the capture and stabilization of N$_2$ by forming B→N π-back bonding and prohibiting H$^+$ binding. In view of this, Zheng *et al.* developed B-doped graphene (BG) by annealing treatment of lyophilized mixtures of boric acid and graphene oxide (GO) with different mass ratios (5:1, 1:10 and 0:1, and the resulting products are denoted as BG-1, BG-2, and G0) under H$_2$/Ar atmosphere. The mixture of boric acid and GO (mass ration of 5:1) treated under Ar atmosphere was denoted as BOG. As expected, BG samples exhibited much-enhanced chemisorption of $N_2$ compared to that of G, and higher B doping leads to improved $N_2$ adsorption. Among these catalysts, BG-1 displayed the best NERR performance in terms of $NH_3$ yield rate (Fig. 8B) and FE of NERR (Fig. 8C) under different potentials. The various N doping types are $BC_3$, $BCO_2$, $BC_2O$ (Fig. 8A). The distal pathway of NERR is dominant for metal-free catalysts. After evaluation of the $N_2$ binding energies, distance, and interaction between *$N_2$ and various types of B atoms, the balance between $N_2$ adsorption and $NH_3$ desorption on the catalysts and formation energies of each intermediate, $BC_3$ was predicted to possess the lowest energy barrier for NERR (Fig. 8D). Based on the experimental measurements and theoretical calculations, the graphene-like BG is a promising metal-free catalyst for NERR.[64]

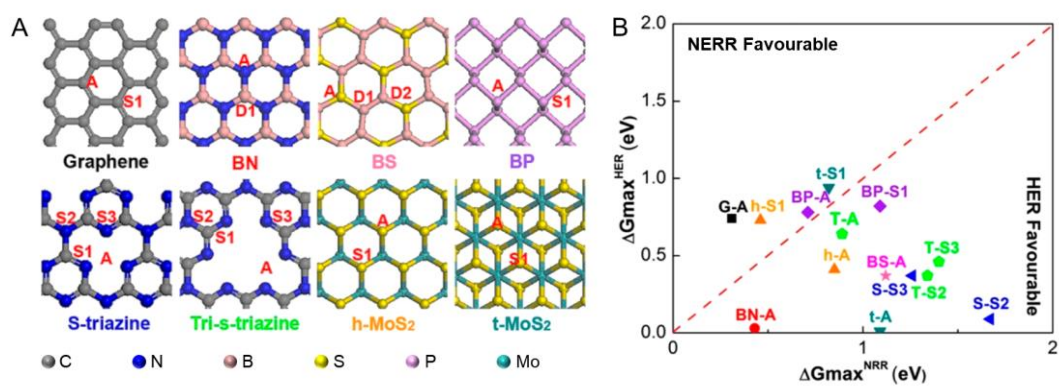

**FIGURE 9.** Proposed structures and calculation Gibbs free energy of various B SACs. (A) Proposed B SACs models by incorporating B atoms on various 2D materials. (B) Maximum Gibbs free energy of various catalysts for HER and NERR. Reproduced with permission.[65] Copyright 2019, American Chemical Society.

On the other hand, B atom has three valence electrons, so its $sp^3$ hybridization contains both empty and occupied orbitals. It has been proven that these B atoms can bind $N_2$ *via* a side-on configuration according to the compatibility of the symmetry of the orbitals.[37] Although several B-based catalysts have been developed, the relationship between the states (chemical environment) of B and NERR performance is poorly known. Sun *et al.* performed a DFT calculation on various B-based SACs for NERR. Unlike metal-based SACs that require empty *d* orbitals to accept the lone-pair electrons of $N_2$, B→N π back bonding could be formed between $N_2$ and B-based SACs, which facilities the adsorption of $N_2$ and weakens the bond energy of $N_2$, therefore lowering the $N_2$ activation barrier. Three states of B, including adsorbed, substituted and lattice B atoms on eight common substrates, and a total of 21 catalysts have been studied (Fig. 9A). The stability simulation of the catalysts including adsorption, cohesive and formation energies indicate that the proposed single B-based structures are stable and can be synthesized. Consistent with the previous report,[64] this research also reveals that the coexistence of occupied and unoccupied states is necessary for B-based NERR catalysts. In addition, the performance of the catalysts for NERR is highly associated with the charge transfer between B and the substrates. Finally, single B atoms substituted into $MoS_2$ and stabilized by graphene are calculated to be the most promising NERR catalysts, presenting superb energy efficiency toward NERR and excellent selectivity against HER (Fig. 9B).[65]

**4.2.2 Others**

In addition, NC, O-doped graphene (OG) and S-doped carbon nanosphere (S/CNS) have been developed for NERR. Derived from zinc-based zeolitic imidazolate frameworks (ZIF-8), metal-free NC was successfully synthesized by Mukherjee *et al.* through a thermal treatment.[66] As expected, the 3D framework construction of the MOF reserve and the defect-rich nanoporous structure was formed by the removal of Zn during the thermal process. The N doping and the graphitization degree could be regulated by changing the annealing conditions to optimize the electronic and geometric structures of NC. The as-synthesized NC exhibited encouraging activity and stability toward NERR, i.e. a high $NH_3$ yield rate of 3.4 µmol/cm$^2$/h (7.3 at 60 °C) and FE of 10.2% at -0.3 V in ambient conditions. DFT results revealed that the active centers, that is, the moiety consist of carbon vacancy and three pyridinic N atoms embedded in a carbon plane, shows a strong affinity to $N_2$ and favors subsequent N≡N cleavage. Surprisingly, additional Fe doping greatly decreases the NERR activity of the NC catalyst because Fe may block the active center by coordinating with pyridinic N. This work paves a new way for design of efficient NERR catalysts.[66] Similarly, NPC was synthesized from the pyrolysis of ZIF-8 and used as a low cost and metal-free electrocatalyst for NERR. Its high content of pyridinic N and pyrrolic N favors $N_2$ adsorption and subsequent N≡N cleavage, which leads to enhanced kinetics of $NH_3$ synthesis. The alternative mechanism of $N_2$ reduction on NPC was revealed by DFT calculations. This finding provides new insight into constructing cost-effective and efficient electrocatalysts for NERR.[67] On the one hand, a group of NC catalysts exhibits an improved catalytic activity in NERR because of the N-containing active centers. On the other hand, N-containing catalysts undergo decomposition during NERR.[68] Smarter design of the stable N-containing catalysts and more rational/comprehensive testing measurements should be conducted to ensure $NH_3$ is solely produced from the $N_2$ reduction. S/CNS was reported as an efficient catalyst for the reduction of $N_2$ to $NH_3$ in a neutral electrolyte. S/CNS achieves an $NH_3$ yield of 19.07 µg/mg$_{cat}$/h and an FE of 7.47% at -0.7 V, both of which are much higher than those of undoped CNS (3.7 µg/mg$_{cat}$/h and 1.45%). The as-synthesized S/CNS exhibited high structural stability and great electrocatalytic durability for NERR.[69] In addition, the NERR performance of OG has been investigated. DFT results reveal that compared to the C–O group, both the O–C=O and C=O groups have a more significant role for NERR.[70]

**Table 1. Summary of catalytic performances for both SACs and related nanocatalysts on NERR.**

| Catalysts | SA content (wt.%) | Catalyst loading (mg/cm$^2$) | Electrolyte | E (V)[a] | FE (%) | Yield rate of $NH_3$ | ref |
|---|---|---|---|---|---|---|---|
| Au SA/C$_3$N$_4$ | 0.15 | 0.48 | 5 mM H$_2$SO$_4$ | -0.1 | 11.1 | 1.305 mg/mg$_{Au}$/h | [42] |
| Au SA/NPC | 0.205 | 0.6 | 0.1 M HCl | -0.2 | 12.3 | 2.32 µg/cm$^2$/h | [7] |
| Au cluster (0.5 nm) on TiO$_2$ | 1.542 | – | 0.1 M HCl | -0.2 | 8.11 | 21.4 µg/mg$_{cat}$/h | [13] |
| Au flower | – | 0.6 | 0.1 M HCl | -0.2 | 6.05 | 25.57 µg/mg$_{Au}$/h | [71] |
| Ru SA/NC | 0.18 | 0.255 | 50 mM H$_2$SO$_4$ | -0.2 | 29.6 | 120.9 µg/mg$_{cat}$/h<br>30.84 µg/cm$^2$/h | [47] |
| Ru SA/NPC | 0.1 | – | 0.1 M HCl | -0.21 | 15; 21[c] | 3.665 mg/mg$_{Ru}$/h | [10] |
| Ru NPs | – | 1.7 | 0.01 M HCl | 0.01<br>-0.1 | 5.4<br>– | –<br>0.55 µg/cm$^2$/h | [72] |
| Mo SA/NPC | 9.54 | 1 | 0.1 M KOH | -0.3 | 14.6±1.6 | 34.0±3.6 mg/mg$_{cat}$/h | [8] |
| | | | 0.1 M HCl | -0.25 | 6.7±0.3 | 31.5±1.2 mg/mg$_{cat}$/h | |
| MoS$_2$ | – | – | 0.1 M Na$_2$SO$_4$ | -0.5 | 1.17 | 4.95 µg/cm$^2$/h | [73] |
| Fe SA/NC | 4.2 | 1 | 0.1 M PBS | -0.4 | 18.6±0.8 | 62.9±2.7 µg/mg$_{cat}$/h | [52] |
| Fe SA/NC | 1.09 | 1 | 0.1 M KOH | 0 | 56.55 | 7.48 µg/mg$_{cat}$/h | [9] |
| Fe SA/NC | 1.2 | 1.71 | 1 M NaOH | -0.05 | 4.51 | 0.95 µg/cm$^2$/h | [74] |
| Fe-N/C-CNT | 0.5 | 0.5 | 0.1 M KOH | -0.2 | 9.28 | 34.83 µg/mg$_{cat}$/h | [75] |
| Fe doped TiO$_2$ | 2.8[d] | 0.1 | 0.5 M LiClO$_4$ | -0.4 | 25.6 | 25.47 µg/mg$_{cat}$/h | [76] |
| Fe/Fe$_3$O$_4$ | – | 0.3 | 0.1 M PBS | -0.3 | 8.29 | 0.19 μg/cm$^2$/h | [77] |
| Y SA/NC | 0.38 | 1.6 | 0.1 M HCl | -0.1 | 12.1 | 23.2 μg/cm$^2$/h | [58] |
| Sc SA/NC | 1.24 | 1.6 | 0.1 M HCl | -0.1 | 11.2 | 20.4μg/cm$^2$/h | |
| B SA/G | 6.2 | 0.178 | 50 mM H$_2$SO$_4$ | -0.5 | 10.8 | 9.8 μg/cm$^2$/h | [64] |
| S-doped CNS | 1.21 | 0.1 | 0.1 M Na$_2$SO$_4$ | -0.7 | 7.47 | 19.07 μg/mg$_{cat}$/h | [69] |
| CNS | – | | | | 1.45 | 3.7 μg/mg$_{cat}$/h | |
| O-doped G | – | – | 0.1 M HCl | -0.55 | – | 21.3 μg/mg$_{cat}$/h | [70] |
| | | | | -0.45 | 12.6 | – | |
| NC | – | 0.8 | 0.1 M KOH | -0.3 | 10.2 | 3.4 μmol/cm$^2$/h | [66] |
| NPC | 13.6 (at%) | 0.6 | 50 mM H$_2$SO$_4$ | -0.9 | 1.42 | 1.40 mmol/g$_{cat}$/h | [67] |

CNT = carbon nanotube. $^a$*vs*. RHE; $^b$synthesized from N$_2$ and H$_2$; $^c$measured at 0.17 V, $^d$the mass ratio of Fe over Ti.

Most reported SACs and related nanocatalysts for NERR have been summarized in Table 1. It can be found that SACs exhibit much higher FE than that of corresponding traditional nanocatalysts, which is ascribed to the unique topographic and electronic structures of SACs and their consequent effects on suppressing HER. Taking the loading amount of catalysts and the content of active atoms (mainly for metal atoms) into account, the greatly enhanced NH$_3$ yield rate was obtained on SACs due to the highly exposed single atoms and improved intrinsic activities derived from the unsaturated coordination and enhanced interactions between single atoms and substrates. Currently, Ru and Fe SACs are superior to others in terms of both FE and the yield rate of NH$_3$. It should be noted that nonmetal-based SACs present substantial NERR performance because nonmetallic atoms could tailor the electronic structure of carbon substrates.

## 5. Summary and perspectives

NERR has been regarded as a promising alternative to the traditional Haber-Bosch process. An increasing number of works aiming at synthesizing novel electrocatalysts with excellent NERR performance have been published and some encouraging results have been obtained. However, most of the catalysts are suffering from the poor FE, low NH$_3$ yield rate and poor selectivity against HER due to the low affinity of N$_2$ on the catalysts, and high overpotential and high energy barrier raised from the intrinsic nature of the catalysts. Fortunately, SACs have emerged as promising NERR electrocatalysts because of their unique structural and electronic characteristics and high atom efficiency. Enormous progress for NERR has been achieved through rational design and the ingenious synthesis of novel SACs including precious metal-, nonprecious metal-, and nonmetal-based SACs. Nevertheless, considerable scientific and technical challenges still exist to promote the wide commercialization of NERR. First of all, the performance of the current SACs for NERR is limited in terms of low activity and selectivity. For example, some catalysts exhibit excellent NERR performance at a relatively high potential. However, at an increased negative potential, all of the catalysts experience dramatically decreased FE and $NH_3$ yield rate because of the thermodynamically favored H over $N_2$ adsorption on catalysts, resulting in the limitation of $N_2$ transport and predominance of HER.[7] Moreover, the mechanism of NERR remains an open debate because of the structural complexity and diversity of the typical catalysts, which therefore constitutes obstacles for reasonable design and the development of advanced NERR catalysts.[15] Therefore, it is urgent and challenging to synthesize novel SACs with efficient and stable NERR activity.

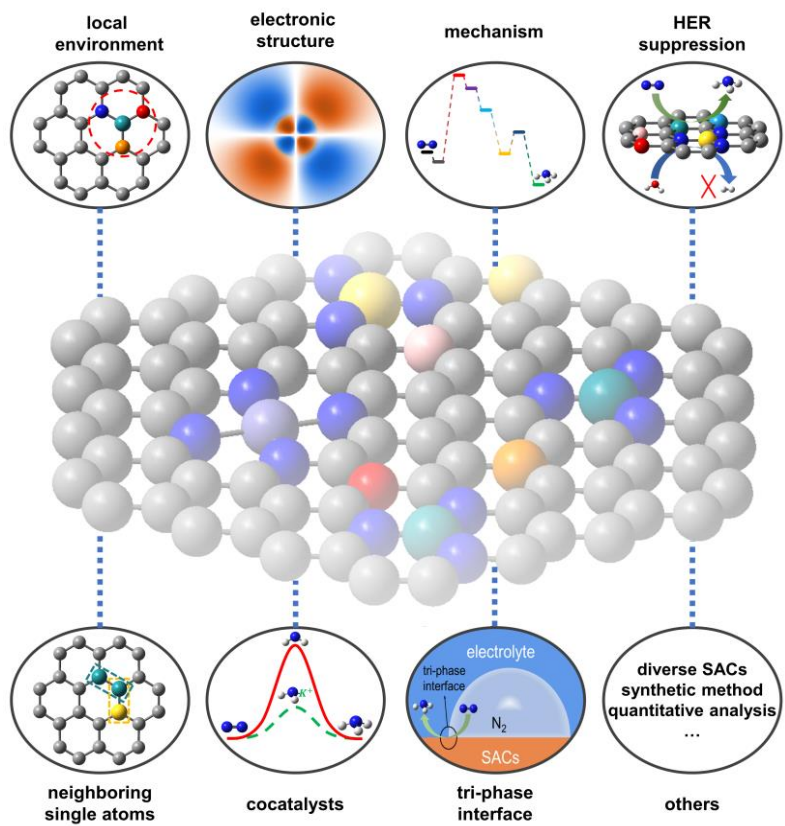

**FIGURE 10**. The perspectives on the future design of SACs and the development of NERR.

To further promote the field development, to our best knowledge, the following strategies could be the effective pathways to design and fabricate adequate and efficient NERR SACs (Fig. 10). (1) NERR involves multiple electron and proton transfer steps and is a complicated process. It is essential to fully understand the structure-performance relationship and catalytic mechanism of NERR at the atomic level for the rational design of the catalysts.[24] More SACs of various single atoms and supports couples should be proposed to expand the library of catalysts for the search of outstanding candidate SACs, which in turn contributes to revealing the mechanism and guiding the design of the SACs. (2) The local environment of the SACs could be regulated to suppress the competing HER. For example, hydrophobic polymers with porous structures were incorporated into the catalyst to obstruct $H_2O$ adsorption on the catalyst surface and thus improve the $NH_3$ yield rate and FE.[78] It is preferred to build an aerophilic-hydrophilic interface between the electrolyte solution and catalysts to simultaneously facilitate the diffusion of $N_2$ and active proton and impede water absorption on the surface of the catalyst surface.[78] (3) Rigorous protocols for precise measurements of $NH_3$ exclusively from electroreduction of $N_2$ are highly desired.[79] The false positives of $NH_3$ from both out-system (ammonia or NOx present in the air, rubber gloves, and human breath) and intrasystem (catalyst decomposition, and N-containing compounds in the feed gas, electrocatalysts, and membrane) would lead to serious questions on the accuracy and reliability of the reported results and inaccurate evaluation of the catalysts.[80,81] (4) More ingenious synthetic methods of the catalysts with desired structures should be explored. A large number of SACs have been predicted as excellent catalysts for NERR, however, few of them have been synthesized experimentally because of the difference in microstructures between ideal models and real samples. (5) Catalysts with neighboring single atoms (both homogeneous and heterogeneous) should be explored to make full use of the synergetic interactions between neighboring coupled single atoms to boost NERR.[82] (6) The effect of cocatalysts should not be ignored. For example, potassium cations ($K^+$) were found to improve NERR and simultaneously suppress HER activity of Bismuth-based catalyst by stabilizing the key $N_2$-reduction intermediates, regulating $H^+$ diffusion and suppressing $H^+$ reduction in aqueous solutions, resulting in an enhanced NERR performance (FE of 66%).[83] Likewise, Li-mediated NERR was achieved because at the given potential $Li^+$ could be reduced to Li, which spontaneously reacts with $N_2$ to form $Li_3N$, which was then protonated to form $NH_3$ when using ethanol as a proton carrier.[84] Moreover, $Li^+$ was demonstrated to retard the HER process because of the formation of active O-Li+ site, and afford a larger potential window for NERR with higher selectivity.[85] (7) Last but not least, exploring novel electrolyte systems with increased $N_2$ solubility is highly desired because the FE of NERR is greatly limited by the low solubility of $N_2$ in aqueous solutions especially at higher current densities.[86] Alternatively, developing air-liquid-solid tri-phase interfaces might promote the NERR as it has been designed to greatly improve the contact area between the gas and catalysts.[87,88] In addition, electrode with gas-diffusion layer has been reported to accelerate the electrocatalysis by increasing the local concentration of gaseous reactant and boosting the diffusion of the production from the active sites.[89] These abovementioned issues should be taken into account to fully understand the $N_2$ reduction process, improve the NERR performance of SACs, and develop energy-efficient $NH_3$ electrochemical synthesis strategies.

## Acknowledgments

J.T. thanks Prof. Dongyuan Zhao at Fudan University for discussing the organization of the article, valuable suggestions and comments on the manuscript. Z.Z., X.Z., and Y. Z. acknowledge the Natural Scientific Foundation of China (21804074), the Natural Scientific Foundation of Shandong Province (ZR2018BB051) and the Open Funds of the State Key Laboratory of Electroanalytical Chemistry (SKLEAC201905). J.C. acknowledges the University of California, Los Angeles for the startup support.